\documentclass[lettersize,journal]{IEEEtran}
\usepackage{amsmath,amsfonts}
\usepackage[noend]{algpseudocode}
\usepackage{algorithmicx,algorithm}
\usepackage{array}
\usepackage[caption=false,font=normalsize,labelfont=sf,textfont=sf]{subfig}
\usepackage{textcomp}
\usepackage{stfloats}
\usepackage{url}
\usepackage{verbatim}
\usepackage{graphicx}
\usepackage{cite}
\usepackage{amssymb}
\newcommand*{\QEDA}{\hfill\ensuremath{\blacksquare}}

\usepackage{color}
\usepackage{multicol}
\usepackage{bm}
\usepackage{latexsym}
\usepackage{float}
\usepackage{exscale}
\usepackage{relsize}
\usepackage{cases}
\usepackage{algpseudocode}
\usepackage{booktabs}
\usepackage{epsfig}
\usepackage{epstopdf}
\usepackage{breqn}
\usepackage{enumerate}
\usepackage{multirow}
\usepackage[utf8]{inputenc}
\usepackage{amsthm}
\usepackage{extarrows}
\UseRawInputEncoding
\begin{document}

\title{Active RIS-Aided Massive MIMO  Uplink Systems with Low-Resolution ADCs}

\author{Zhangjie Peng, Zecheng Lu, Xue Liu, Cunhua Pan, \IEEEmembership{Senior Member,~IEEE},
and Jiangzhou Wang, 
\IEEEmembership{Fellow,~IEEE}       
\thanks{Z. Peng, Z. Lu and X. Liu are with the College of Information, Mechanical and Electrical
	Engineering, Shanghai Normal University, Shanghai 200234, China.(e-mail:
	pengzhangjie@shnu.edu.cn; 1000511822@smail.shnu.edu.cn; 1000494962@smail.shnu.edu.cn).}

\thanks{C. Pan is with the National Mobile Communications
	Research Laboratory, Southeast University, Nanjing 210096, China.
	(e-mail: cpan@seu.edu.cn).
}
\thanks{J. Wang is with the School of Engineering, University of Kent, CT2 7NT Canterbury, U.K. (e-mail: j.z.wang@kent.ac.uk).
}
}

\markboth{}%
{Shell \MakeLowercase{\textit{et al.}}: A Sample Article Using IEEEtran.cls for IEEE Journals}


\maketitle
\newtheorem{lemma}{Lemma}
\newtheorem{theorem}{Theorem}
\newtheorem{remark}{Remark}
\newtheorem{corollary}{Corollary}
\newtheorem{proposition}{Proposition}\vspace{-0.1cm}
\begin{abstract}
This letter considers an active reconfigurable intelligent surface (RIS)-aided multi-user uplink massive multiple-input multiple-output (MIMO) system with low-resolution analog-to-digital converters (ADCs). The letter derives the closed-form approximate expression for the sum achievable rate (AR), where the
maximum ratio combination (MRC) processing and low-resolution ADCs are applied at the base station. The system performance is analyzed, and a genetic algorithm (GA)-based method is proposed to optimize the RIS's phase shifts for enhancing the system
performance. Numerical results verify the accuracy of the
derivations, and demonstrate that the active RIS has an evident performance gain over the passive RIS.
\end{abstract}

\begin{IEEEkeywords}
Reconfigurable intelligent surface (RIS), active RIS, uplink achievable rate, massive MIMO,  low-resolution ADCs.
\end{IEEEkeywords}

\section{Introduction}
Thanks to the capability of reconfiguring the radio propagation environment, reconfigurable intelligent surface (RIS) consisting of numerous reflecting elements has been considered as a breakthrough technology of 6th generation mobile networks\cite{9847080}, \cite{9459505}. Specifically, RIS has low deployment costs and energy consumption while improving the performance and coverage of communication systems \cite{8811733}.

However, most of the existing  contributions considered passive RIS. In fact, passive RIS suffers from  ``multiplicative fading'', which leads to severe signal attenuation  and limits the system performance\cite{8888223}. To solve this problem, active RIS  equipped with active reflection-type amplifiers was proposed and regarded as a promising solution. This technology not only enables the adjustment of phase shifts but also amplifies the received signal at the cost of additional hardware power consumption\cite{9377648}, \cite{9530750}. Recent contribution indicates that  active RIS has a notable improvement over passive RIS under the same total network power consumption including both the transmit power and the hardware power consumption\cite{9734027}. The researchers have explored the application of active RIS in various scenarios, such as  multi-pair device-to-device (D2D) communications systems\cite{9840889}, mobile edge computing systems \cite{9878164} and multiuser communication systems with hardware impairments \cite{10452289}. 

Recently, researches have integrated RIS into massive multiple-input multiple-output (MIMO)  communication systems, since it can greatly improve system performance \cite{9743440}. With high-resolution digital-to-analog converters (DACs) or analog-to-digital converters (ADCs) equipped on each antenna, massive MIMO systems require substantial power consumption and cost expenditures. A promising solution for tackling this issues is to apply low-resolution ADCs/DACs in the systems because of its cost effectiveness \cite{7307134}. The authors of \cite{9833357} and \cite{9483943} investigated low-resolution ADCs/DACs in passive RIS-aided massive MIMO systems and analyzed the system capacity.

However, as far as we are aware, the active RIS-aided multi-user uplink massive MIMO system with low-resolution ADCs has not been studied. The specific contributions of this letter are summarized as: 1) For an active RIS-aided multi-user uplink massive MIMO system with low-resolution ADCs, the closed-form approximate expression of the sum achievable rate (AR) is derived; 2) A genetic algorithm (GA)-based method is proposed to maximize the sum AR through the RIS's phase shifts optimization; 3) Extensive numerical results verify the accuracy of the derivations, and show the
performance advantage of the active RIS and the rationality
of applying low-resolution ADCs.

\begin{figure}
	\centering
	\setlength{\abovecaptionskip}{-1ex}
	\includegraphics[scale=0.42]{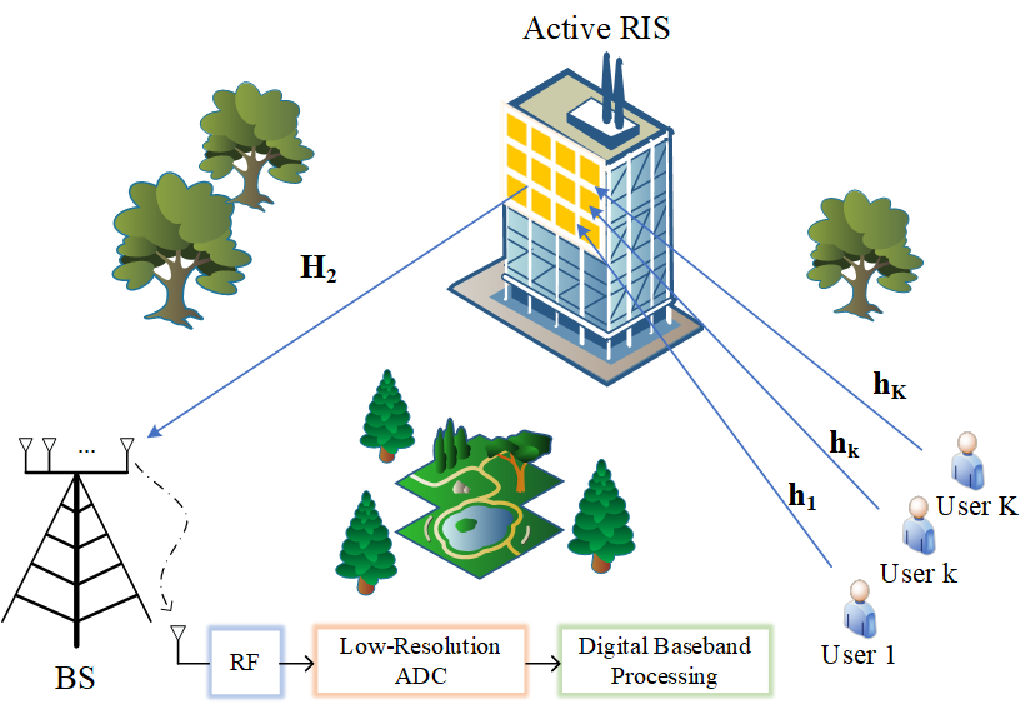}
	\caption{Active RIS-aided  massive MIMO uplink system with low-resolution ADCs.}
	\label{Figsysmodel}
	\vspace{-3.2ex}
\end{figure}

\section{System Model}\label{SYSTEM MODEL AND PROBLEM FORMULATION}

An active RIS-aided multi-user uplink massive MIMO system with low-resolution ADCs is depicted in Fig. \ref{Figsysmodel}. The direct communication links of this system are  hindered owing to the obstacles. Hence, an active RIS with $ N $ reflecting elements is used to assist the communications between $K$ single-antenna users and a base station (BS) that is composed of $M$ antennas.
\vspace{-2ex}
\subsection{Channel Model}
 All channels of this system follow the Rician fading. The users-RIS channel matrix ${\mathbf{H}_{1}} \in {{\mathbb C}^{{N} \times {K}}}$, the $ k $-th user-RIS channel vector ${\mathbf{h}_{k}} \in {{\mathbb C}^{{N} \times {1}}}$ for $ k\in{\cal{S}} \triangleq\left\{1,2,...,K\right\} $ and the RIS-BS channel matrix ${\mathbf{H}_{2}} \in {{\mathbb C}^{{M} \times {N}}}$ are respectively expressed as
\vspace{-0.5ex}
\begin{equation}
	{\mathbf{H}_{\rm 1}}=\left[\mathbf{h}_{\rm 1},...,\mathbf{h}_{k},...,\mathbf{h}_{K}\right],\quad\quad\quad\quad\quad\quad\quad
\end{equation}
\begin{equation}
	\mathbf{h}_{k}=\sqrt{\alpha_{k}}\left(\sqrt{\frac{\varepsilon_{k}}{\varepsilon_{k} +1}}\bar{\mathbf{h}}_{k}+\sqrt{\frac{1}{\varepsilon_{k} +1}}\widetilde{\mathbf{h}}_{k}\right)\!,
\end{equation}
\begin{equation}
	\!\mathbf{H}_{\rm 2}=\sqrt{\beta}\left(\sqrt{\frac{\delta}{\delta +1}}\bar{\mathbf{H}}_{\rm 2}+\sqrt{\frac{1}{\delta +1}}\widetilde{\mathbf{H}}_{\rm 2}\right),\ 
\end{equation}
where $ \alpha_{k} $ and $ \beta $ stand for the corresponding large-scale fading coefficients.
Vector $ \bar{\mathbf{h}}_{k} $ and matrix $ \bar{\mathbf{H}}_{\rm 2} $ denote the light-of-sight (LoS) parts. The parts of non-line-of-sight (NLoS) are respectively denoted by vector $ \widetilde{\mathbf{h}}_{k} $ and matrix $ \widetilde{\mathbf{H}}_{\rm 2} $ with independently and identically distributed
(i.i.d) elements following the distribution of $ \mathcal{CN}(0,1) $. $ \varepsilon_{k} $ and $ \delta $ represent the Rician factors of $ \mathbf{h}_{k} $ and $ \mathbf{H}_{2} $, respectively.

Assuming that uniform square planar arrays are adopted at the BS and the RIS, $\bar{\mathbf{h}}_{k} $ and $ \bar{\mathbf{H}}_{\rm 2} $ can be respectively given by
\begin{equation}\label{4}
	\!\bar{\mathbf{h}}_{k}=\mathbf{a}_N(\phi_{k\mathrm{r}}^a,\phi_{k\mathrm{r}}^e),\quad\quad\quad\quad\quad\ 
\end{equation}
\vspace{-3ex}
\begin{equation}\label{5}
	\bar{\mathbf{H}}_{\rm 2}=\mathbf{a}_M(\psi_{\mathrm{R}k}^a,\psi_{\mathrm{R}k}^e)\mathbf{a}_N^H(\phi_{\mathrm{t}}^a,\phi_{\mathrm{t}}^e),
\end{equation}
where $ \phi_{kr}^a $ and $ \phi_{kr}^e $ are the azimuth and elevation angles of arrival (AOA) from the $ k $-th user to the active RIS, respectively. Then, $ \phi_{t}^a $ and $ \phi_{t}^e $ stand for the azimuth and elevation angles of departure (AOD) at the BS from the active RIS,  respectively. $ \psi_{Rk}^a $ and $ \psi_{Rk}^e $ represent the corresponding azimuth and elevation AOA at the BS from the active RIS. By denoting $ \lambda $ as the wavelength, the $ L $ dimensional array response vector $ \mathbf{a}_L(\varphi^a,\varphi^e) $ is given by 
\begin{align}\label{6}
	\!\!\mathbf{a}_L(\varphi^a,\varphi^e)\!=&\big[1,...,e^{j2\pi\!\frac{d}{\lambda}(x\!\sin{\varphi^e}\!\sin{\varphi^a}+y\!\cos{\varphi^e})}, ...\notag\\ &,e^{j2\pi\!\frac{d}{\lambda}((\sqrt{L}-1)\!\sin{\varphi^e}\!\sin{\varphi^a}+(\sqrt{L}-1)\!\cos{\varphi^e})}\big]^T\!,
\end{align}
where $ x=\lfloor \left( l-1 \right) /\sqrt{L} \rfloor  $, $ y=\left( n-1 \right) \mathrm{mod}\sqrt{L} $ and $ d $ is the elements spacing.
\vspace{-2ex}
\subsection{Signal Transmission Model}
The amplified signal at the active RIS can be represented as
\vspace{-1ex}
\begin{align}\label{y_R}
	\mathbf{y}_{\mathrm{R}}= \mathbf{A}\mathbf{\Phi }{\mathbf{H}_1}\mathbf{P}\mathbf{x} + \mathbf{A}\mathbf{\Phi }\mathbf{v},
\end{align}
where $ \mathbf{A}\triangleq \mathrm{diag}\left\{\eta_1,...,\eta_n,...,\eta_N\right\}\in \mathbb{R}^{N\times N} $ is the amplification factor matrix and each element in the matrix is larger than 1. $ \mathbf{P}\triangleq \mathrm{diag}\left\{\sqrt{p_1},...,\sqrt{p_k},...,\sqrt{p_K}\right\}\in \mathbb{R}^{K\times K} $, where $ p_k $ denotes the transmit power of the $ k $-th user. The
reflecting coefficient matrix of the active RIS is denoted by $\mathbf{\Phi }=\mathrm{diag}\left\{e^{j\theta_1} ,...,e^{j\theta_N} \right\}\in {{\mathbb C}^{{N} \times {N}}} $, and $ \theta_n \in [0,2\pi)$ stands for the phase shift of element $ n $. $ \mathbf{x}\triangleq[x_1,x_2,...,x_K]^T $ is the transmit symbol with $ \mathbb{E}\small\{\left\vert x_k \right\vert^2\small\}=1 $. 
Owing to the application of the active components, we express $ \mathbf{A}\boldsymbol{\Phi}\mathbf{v} $ as the dynamic noise \cite{zhang2021active}, where $ \mathbf{v}\sim \mathcal{CN}(0_N,\sigma_{\mathrm{v}}^2 \mathbf{I}_N)  $.
\begin{figure*}[hb]
	\vspace{-2ex}
	\hrulefill
	\setcounter{equation}{14}
	\begin{align}\label{15}
		R_k\!=\!\mathbb{E}\left\{\!\log_{2}\! \left(\!1\!+\!\frac{{p_k}{\alpha^2}{\lVert \mathbf{g}_k \rVert}^4}{{\alpha^2\!\!\sum\limits_{i = 1,i \ne k}^K {{p_i}{\left\vert {\mathbf{g}_k^H}{\mathbf{g}_i} \right\vert}^2}}\!+\!{\eta^2}{\alpha^2}{\sigma_{\mathrm{v}}^2}{\lVert {\mathbf{g}_k^H}{\mathbf{H}_2}{\mathbf{\Phi}} \rVert}^2\!+\!{\alpha^2}{\sigma_n^2}{\lVert \mathbf{g}_k \rVert}^2\!\!+\!\alpha\! \left( 1\!-\!\alpha \right) \mathbf{g}_{k}^{H}\mathrm{diag}\left(\!p_k\mathbf{GG}^H\!\!+\!\sigma _{n}^{2}\mathbf{I}_M\! \right)\! \mathbf{g}_k}\!\right)\!\right\}.
	\end{align}
    \setcounter{equation}{16}
    \vspace{-1.3ex}
	\hrulefill
    \begin{align}\label{17}
    	\Gamma _1=\eta ^4Mu_k ^2
    	\big\{ M\delta ^2\varepsilon _{k}^{2}\left| f_k\left( \mathbf{\Phi } \right) \right|^4+2\delta \varepsilon _k\left| f_k\left( \mathbf{\Phi } \right) \right|^2\times
    	\left( 2MN+MN\varepsilon _k+MN+2M+N\varepsilon _k+N+2 \right)\quad\quad\quad\quad\quad\quad\quad\quad\ \  \notag 
    \end{align}
\vspace{-5ex}
\begin{align}
    	\quad\quad+MN^2\!\left( 2\delta ^2\!+\!\varepsilon _{k}^{2}\!+\!2\delta \varepsilon _k\!+\!2\delta \!+\!2\varepsilon _k\!+\!1 \right) 
    	\!+\!N^2\left( \varepsilon _{k}^{2}+2\delta \varepsilon _k\!+\!2\delta \!+\!2\varepsilon _k\!+\!1 \right) 
    	\!+\!MN\!\left( 2\delta \!+\!2\varepsilon _k\!+\!1 \right) \!+\!N\left( 2\delta \!+\!2\varepsilon _k\!+\!1 \right) \big\}, \ 
    \end{align}
\vspace{-4.5ex}
\begin{align}\label{18}
	\Gamma _2=\eta ^4Mu_k^2 u_i^2
	\big\{ M\delta ^2\varepsilon _k\varepsilon _i\left| f_k\!\left(\! \mathbf{\Phi } \!\right) \right|^2\left| f_i\!\left(\! \mathbf{\Phi } \!\right) \right|^2
	\!+\!\delta \varepsilon _k\left| f_k\!\left(\! \mathbf{\Phi } \!\right) \right|^2\!\left( \delta MN+N\varepsilon _i+N\!+\!2M \right) 
	\!+\!\delta \varepsilon _i\left| f_i\!\left(\! \mathbf{\Phi } \!\right) \right|^2\!
	\left( \delta MN\!+\!N\varepsilon _k\!+\!N\!+\!2M \right) \!\!\quad\notag
\end{align}
\vspace{-5ex}
\begin{align}
	\quad\quad+N^2\!\!\left( \!M\delta ^2\!\!+\!\delta\! \left( \varepsilon _k\!+\!\varepsilon _i\!+\!2 \right) \!\!+\!\!\left( \varepsilon _i\!+\!1 \right) \left( \varepsilon _k\!+\!1 \right)\! \right) 
	\!\!+\!\!M\!N\!\left( \!2\delta \!+\!\varepsilon _k\!+\!\varepsilon _i\!+\!1\! \right) \!\!+\!\!M\varepsilon _k\varepsilon _i\!\left| \mathrm{\bar{\mathbf{h}}}_{k}^{H}\!\mathrm{\bar{\mathbf{h}}}_i \right|^2
	\!\!+\!2M\!\delta \varepsilon _k\varepsilon _i\mathrm{Re}\!\left\{\! f_{k}^{H}\!\!\left(\! \mathbf{\Phi } \!\right)\! f_i\!\left(\! \mathbf{\Phi } \!\right)\! \mathrm{\bar{\mathbf{h}}}_{k}^{H}\!\mathrm{\bar{\mathbf{h}}}_i \!\right\} \!\!\big\},\!\!\!
\end{align}
\vspace{-4.5ex}
\begin{align}\label{19}
\Gamma _3=\frac{\eta ^2M^2\beta u_k}{\left( \delta +1 \right)}\!\left(\! \delta \varepsilon _k\left( 2\!+\!\delta N \right) \left| f_k\left( \mathbf{\Phi } \right) \right|^2+2N\delta \!+\!N^2\delta ^2+N\varepsilon _k+N \right) \!+\!\eta ^2MN\beta u_k\left( \delta \varepsilon _k\left| f_k\left( \mathbf{\Phi } \right) \right|^2\!+\!N\delta \!+\!N\varepsilon _k+N \right)\!,\ \ 
\end{align}
\vspace{-4ex}
\begin{align}\label{20}
	\!\!\Gamma _4=\eta ^2Mu_k
	\small( \delta \varepsilon _k\left| f_k\left( \mathbf{\Phi } \right) \right|^2+\delta N+\varepsilon _kN+N \small),\quad\quad\quad\quad\quad\quad\quad\quad\quad\quad\quad\quad\quad\quad\quad\quad\quad\quad\quad\quad\quad\quad\quad\quad\quad\quad\quad\quad\quad\quad\quad\quad
\end{align}
\vspace{-4.5ex}
\begin{align}\label{21}
	\!\Gamma _5=p_k\eta ^4Mu_{k}^{2}\big\{ \left( \delta \varepsilon _k\left| f_k\left( \mathbf{\Phi } \right) \right|^2 \right) ^2\!+\!4\delta \varepsilon _k\left| f_k\left( \mathbf{\Phi } \right) \right|^2\left( N\left( \delta +\varepsilon _k+1 \right) +2 \right) +2N^2\left( \delta +\varepsilon _k+1 \right) ^2+2N\left( 2\delta +2\varepsilon _k+1 \right) \big\}\quad\ \  \notag
\end{align}
\vspace{-4.5ex}
\begin{align}
	\quad\quad\!+\eta ^2\!\sigma _{n}^{2}\!M\!u_k\!\left(\! \delta \varepsilon _k\!\left| f_k\!\left( \!\mathbf{\Phi } \!\right) \right|^2\!\!\!+\!\!N\!\left(\delta \!+\!\varepsilon _k\!+\!1\right) \!\right) 
	\!\!+\!p_k\eta ^4\!M\!\sum\nolimits_{i=1\!,i\ne k}^K{}\!\!\big\{ \!u_k\!u_i\!\left(\! \delta \varepsilon _k\!\left| f_k\!\left( \!\mathbf{\Phi } \!\right) \right|^2\!\!\!+\!\!N\!\left(\delta \!+\!\varepsilon _k\!+\!1\right)\! \right) \!\!\left(\! \delta \varepsilon _i\!\left| f_i\!\left(\! \mathbf{\Phi }\! \right) \right|^2\!\!\!+\!\!N\!\left(\delta \!+\!\varepsilon _i\!+\!1\right)\!\right)\notag
\end{align}
\vspace{-4.5ex}
\begin{align}
		\quad\quad\!\!\!+2\delta u_ku_i \left( \varepsilon _k\varepsilon _i\mathrm{Re}\left\{ f_{k}^{H}\!\left( \mathbf{\Phi } \right) f_i \!\left( \mathbf{\Phi } \right) \mathbf{\bar{h}}_{k}^{H}\mathbf{\bar{h}}_i \right\} +\varepsilon _k\left| f_k\!\left( \mathbf{\Phi } \right) \right|^2+\varepsilon _i\left| f_i \!\left( \mathbf{\Phi } \right) \right|^2+N \right) \big\},\quad\quad\quad\quad\quad\quad\quad\quad\quad\quad\quad\quad\quad\ \ 
\end{align}

\ \ where $ f_{\mathrm{s}}\left( \mathbf{\Phi } \right) \triangleq \mathbf{a}_{N}^{H}\left( \phi _{t}^{a},\phi _{t}^{a} \right) \mathbf{\Phi \bar{\mathbf{h}}}_{\mathrm{s}}  $ and $ u_{\mathrm{s}}\triangleq \frac{\beta \alpha _{\mathrm{s}}}{\left( \delta +1 \right) \left( \varepsilon _{\mathrm{s}}+1 \right)} $, 
 $ {\mathrm{s}}\in{\left\{k,i\right\}} $ .

	\setcounter{equation}{7}
\end{figure*}
\begin{table}[t]
	\setlength{\abovecaptionskip}{0.1cm}  
	\setlength{\belowcaptionskip}{-0.2cm} 
	\tabcolsep=0.2cm
	\caption{$ \rho $ FOR DIFFERENT ADC QUANTIZATION BITS $ b $}
	\vspace{0.2cm}
	\label{table1}
	\vspace{-0.4cm}
	\centering
	\begin{tabular}{ccccccc}
		\toprule
	$ b\! $& 1 & 2 & 3 & 4 & 5 & $\geqslant$ 6 \\
		\midrule
		$ \rho\! $& 0.3634 & 0.1175 & 0.03454 & 0.009479 & 0.002499 & $ \dfrac{\pi\sqrt{3}}{2}\!\cdot\!2^{-2b}\! $\\
		\bottomrule
	\end{tabular}
	
\end{table}
\vspace{0ex}

We assume that the values of each element in matrix $ \mathbf{A} $ are equal, i.e., $ \eta_n\triangleq\eta $. Thus, the reflecting signal power of the active RIS $ P_{\mathrm{A}}\triangleq\mathbb{E}\small\{ {{{\left\| {{{\mathbf{y}}_{\mathrm{R}}}} \right\|}^2}} \small\} $ is  calculated as
\vspace{-1ex}
\begin{align}\label{p_ris}
	&\mathbb{E}\left\{ {{{\left\| {{{\mathbf{y}}_{\mathrm{R}}}} \right\|}^2}} \right\} = \mathbb{E} \left\{ \left\| \mathbf{A\Phi H}_1\mathbf{Px}+\mathbf{A\Phi v} \right\| ^2 \right\}\notag\\ 
	&\!=\!\mathrm{Tr}\left( \mathbb{E} \left\{ \left( \mathbf{A\Phi H}_1\mathbf{Px}+\mathbf{A\Phi v} \right) \left( \mathbf{A\Phi H}_1\mathbf{Px}+\mathbf{A\Phi v} \right) ^H \right\} \right)\notag\\ 
	&\!=\!\eta ^2\!\!\left(\!\mathrm{Tr}\!\left(\! \mathbb{E}\! \big\{\! \mathbf{\Phi H}_1\!\mathbf{P}\mathbb{E}\! \left\{\! \mathbf{xx}^H\! \right\}\!\! \mathbf{P}^H\!\mathbf{H}_{1}^{H}\!\mathbf{\Phi }^H\! \big\} \!\right) 
	\!\!+\!\!\mathrm{Tr}\!\left(\! \mathbb{E} \!\left\{\! \mathbf{\Phi vv}^H\!\mathbf{\Phi }^H\! \right\}\! \right)\!\right)\!\!\notag\\ 
	&\!=\!\eta ^2\mathrm{Tr}\left( \mathbb{E} \left\{ \mathbf{H}_{1}^{H}\mathbf{H}_1\mathbf{PP}^H \right\} \right) +\eta ^2\mathrm{Tr}\left( \mathbb{E} \big\{ \mathbf{vv}^H \big\} \right) \notag\\
	&\!=\!\eta ^2N\big( \sum\nolimits_{k=1}^K{p_k\alpha _k+\sigma _{v}^{2}} \big).
\end{align}

The signal received at the BS is
\begin{align}\label{received signal BS}
	\mathbf{y}=\mathbf{H}_2\mathbf{y}_{\mathrm{R}}+\mathbf{n}=\mathbf{G}\mathbf{P}\mathbf{x} \!+\! \mathbf{H}_2\!\mathbf{A}\mathbf{\Phi }\mathbf{v}\! +\! \mathbf{n}
	,
\end{align}
where $ \mathbf{G}\triangleq\mathbf{H_2}\mathbf{A}\mathbf{\Phi }\mathbf{H_1}\in \mathbb{R}^{M\times K} $ denotes the cascaded channel of the communication links and $ \mathbf{n}\sim \mathcal{CN}(0,\sigma_n^2 \mathbf{I}_M) $ stands for the additive white Gaussian noise (AWGN). 

For reducing the costs of the actual deployment, this letter applies low-resolution ADCs on the basis of the additive quantization noise model (AQNM) at the BS, and the quantized signal is
obtained as\cite{7307134}
\begin{align}\label{y_q}
	\mathbf{y}_{\mathrm{q}}&=\alpha\mathbf{y}+\mathbf{n}_{\mathrm{q}}=\alpha\mathbf{G}\mathbf{P}\mathbf{x} + \alpha\mathbf{H}_2\mathbf{A}\mathbf{\Phi }\mathbf{v} + \alpha\mathbf{n} + \mathbf{n}_{\mathrm{q}},
\end{align}	
where $ \alpha=1-\rho $. Furthermore, $ \rho $ represents the inverse of the signal-to-quantization-noise ratio. The corresponding relationship between $ \rho $ and quantization bit $ b $ is listed in Table I. $ \mathbf{n}_{\mathrm{q}} $ represents the additive Gaussian quantization noise vector, and the covariance of $ \mathbf{n}_{\mathrm{q}} $ is obtained as
\begin{equation}\label{R_nqnq}
	\mathbf{R}_{{\mathbf{n}_{\mathrm{q}}}{\mathbf{n}_{\mathrm{q}}}}=\alpha \left( 1-\alpha \right) \mathrm{diag}\left( \mathbf{GPG}^H+\sigma _{n}^{2}\mathbf{I}_M \right).
\end{equation}

According to \eqref{y_q}, the signal processed by maximal-ratio-combining (MRC) is given by
\begin{align}\label{r_at_BS}
	\!\!\mathbf{r}\!=\!\mathbf{G}\!^H\!\mathbf{y}_{\mathrm{q}}\!=\!\alpha\mathbf{G}\!^H\!\mathbf{G}\mathbf{P}\mathbf{x} \!+\! \alpha\mathbf{G}\!^H\!\mathbf{H}_2\!\mathbf{A}\mathbf{\Phi }\mathbf{v} \!+\! \alpha\mathbf{G}\!^H\!\mathbf{n}\! +\! \mathbf{G}\!^H\!\mathbf{n}_{\mathrm{q}}.
\end{align}
Hence, the signal transmitted by the $ k $-th user can be further written as
\vspace{-2ex}
\begin{align}
	r_k=&\ {\alpha}\sqrt{p_k}{\mathbf{g}_k^H}{\mathbf{g}_k}{x_k}+\alpha\!\!\sum\limits_{i = 1,i \ne k}^K {\sqrt {{p_i}}{\mathbf{g}_k^H}{\mathbf{g}_i}x_i}\nonumber\\
	&+\eta\alpha{\mathbf{g}_k^H}{\mathbf{H}_2}\mathbf{\Phi}\mathbf{v}+\alpha{\mathbf{g}_k^H}{\mathbf{n}}+{\mathbf{g}_k^H}{\mathbf{n}_{\mathrm{q}}},		
\end{align}
where $ \mathbf{g}_k\triangleq\mathbf{H}_2\mathbf{A}\mathbf{\Phi }\mathbf{h}_k=\eta\mathbf{H}_2\mathbf{\Phi }\mathbf{h}_k $. 
Then, the sum AR can be expressed as 
\vspace{-1ex}
\begin{equation}
	R=\sum_{k=1}^K R_k
	,
\end{equation}
where $ R_k $ represents the uplink AR of the $ k $-th user. We give the expression of $ R_k $ in \eqref{15} at the bottom of the next page.
\section{Analysis Of Achievable Uplink Rate}
In this section, the closed-form approximate expression of the uplink AR is derived. Based on the expression, we can analyze the impact of various parameters on system performance. 
\vspace{-1ex}
\begin{theorem}\label{thm_appro}
	For an active RIS-aided massive MIMO system with low-resolution ADCs and MRC processing, the uplink AR of the $k$-th user can be approximated as
\end{theorem}
\vspace{-3ex}
	\setcounter{equation}{15}
\begin{align}\label{16}
	R_k \approx \log _2\!\left(\! 1+\frac{p_k\Gamma _1}{\sum_{i=1,i\ne k}^{K}{p_i\Gamma _2\!+\!\eta ^2\sigma _{\mathrm{v}}^{2}\Gamma _3\!+\!\sigma _{n}^{2}\Gamma _4\!+\!\frac{1-\alpha}{\alpha}\Gamma _5}} \!\right)\!,
\end{align}
where $ \Gamma _1 $, $ \Gamma _2 $, $ \Gamma _3 $, $ \Gamma _4 $ and $ \Gamma _5 $ are respectively given by $ \mathbb{E} \small\{ \left\| \mathbf{g}_k \right\| ^4 \small\} $, $ \mathbb{E} \small\{ \left| \mathbf{g}_{k}^{H}\mathbf{g}_i \right|^2 \small\} $, $ \mathbb{E} \small\{ \left\| \mathbf{g}_{k}^{H}\mathbf{H}_2\mathbf{\Phi } \right\| ^2 \small\} $, $\mathbb{E} \small\{ \left\| \mathbf{g}_k \right\| ^2 \small\}  $ and $ \mathbb{E} \small\{ \mathbf{g}_{k}^{H}\mathrm{diag}\small( p_k\mathbf{GG}^H+\sigma _{A}^{2}\mathbf{I}_M \small) \mathbf{g}_k \small\} $ at the bottom of this page.\vspace{0.5ex}

\ \ \ \emph{Proof:}	See Appendix.\QEDA

According to Theorem \ref{thm_appro}, it indicates that the uplink AR of this system is affected by the number of BS antennas, the number of RIS reflecting elements, the large-scale fading coefficients, the transmit power, the noise power, the dynamic noise power, the Rician factors, the quantization
bit, the AOA, the AOD and the power of active RIS.

Moreover, the power consumption model of the considered system is
\setcounter{equation}{21}
\vspace{-1ex}
\begin{align}
	P_{\mathrm{T}}=P_{\mathrm{t}}+P_{\mathrm{A}}+N\left( P_{\mathrm{SW}}+P_{\mathrm{DC}} \right), 
\end{align}
where $ P_{\mathrm{t}}=\sum_{i=1}^{K}{p_i} $. $ P_{\mathrm{SW}} $ and $ P_{\mathrm{DC}} $ represent the direct current biasing power and the power consumed by the switch and control circuit at each reflecting element,  respectively \cite{9377648}. The active RIS should meet the startup condition $ P_{\mathrm{T}}\geqslant N\left( P_{\mathrm{SW}}+P_{\mathrm{DC}} \right)  $, otherwise $ R_k = 0 $.
\vspace{-1.5ex}
\begin{corollary}\label{c1}
When $ \eta ^2=1 $ and $ \sigma _{v}^{2}=0 $ in \eqref{16}, the uplink AR in the case with passive RIS can be approximated as 
\vspace{-1ex}
\begin{align}\label{22}
	R_{\mathrm{p}} \approx \log _2\!\left(\! 1+\frac{p_k\Gamma _1}{\sum_{i=1,i\ne k}^{K}{p_i\Gamma _2\!+\!\sigma _{n}^{2}\Gamma _4\!+\!\frac{1-\alpha}{\alpha}\Gamma _5}} \!\right)\!.
\end{align}
\end{corollary}
The total power consumption is $ P_{\mathrm{T}}=P_{\mathrm{t}}^{'}+NP_{\mathrm{SW}} $ and the threshold condition is $ P_{\mathrm{T}}\geqslant NP_{\mathrm{SW}} $.
\vspace{-1ex}
\begin{corollary}\label{c2}
With ideal ADCs ($ b\!\rightarrow\!\infty $, $ \alpha\!\rightarrow\!1 $), the uplink AR in \eqref{16} converges to $ R_{k}\rightarrow R_{\mathrm{ideal}}  $, where 
\vspace{-1.5ex}
\begin{align}\label{23}
	R_{\mathrm{ideal}} = \log _2\!\left( 1+\frac{p_k\Gamma _1}{\sum_{i=1,i\ne k}^{K}{p_i\Gamma _2\!+\!\eta ^2\sigma _{v}^{2}\Gamma _3\!+\!\sigma _{n}^{2}\Gamma _4\!}} \right)\!,
\end{align}

\end{corollary}
\vspace{-1ex}
\noindent which means the system has infinite precision. Due to $ b\!\rightarrow\!\infty $, we consider that it is reasonable to ignore the quantization error of ADCs.
\vspace{1.5ex}

  \begin{algorithm}[t]\small
	\renewcommand{\algorithmicrequire}{\textbf{Input:}}
	\renewcommand{\algorithmicensure}{\textbf{Output:}}
	\caption{GA-Based Method}
	\label{alg:1}
	\begin{algorithmic}[1]
		\Require A population $ \rm{P}_1 $ with $ N_{\rm{total}} $ individuals, the iteration number
		$ i $ = 1, the number of the termination iteration times is $ i_T $ , and the termination value is $ f_i $.
		\Ensure maximum fitness function values
		\While {$ i\!<\!i_T $ or $\bar{f}$(average change of the fitness value)$<\! f_i$} 
		\State   Calculate the fitness value $\bar{f}$ according to \eqref{16};
		\State Select $ N_b $ individuals with large fitness function values as elites; 
		\State Select $N_p$ parents  ($N_p\cap N_b =\varnothing $) for crossover to generate $ N_c $ offspring;
		\State Use the remaining individuals for mutation to generate $ N_d $ offspring;
		\State Add $ N_b $, $ N_c $ and $ N_d $ to the next population $ \rm{P}_{i+1} $; $ i=i+1 $.
		\EndWhile 
	\end{algorithmic}  
\end{algorithm}

\vspace{-3ex}
\section{Phase Shifts Optimization}\label{4}
\vspace{-1ex}
In this section, a GA-based method for optimizing phase shifts is proposed to enhance the system performance. The optimization problem is formulated as 
\vspace{-1ex}
\begin{align}
\max_{\mathbf{\Phi }} \sum\nolimits_{k=1}^K{R_k}\quad\quad\ \ 
\end{align}
\vspace{-4ex}
\begin{align}\label{24}
\mathrm{s}.\mathrm{t}. \ \theta _n\in \left[ 0,2\pi \right) ,\forall n,
\end{align}
where $ R_k $ is given by \eqref{16} in Theorem \ref{thm_appro}. The specific content of the GA-based method is displayed in
Algorithm 1 at the top of this page.

\section{Numerical Results}
In this part, we analyze the impacts of the number of BS antennas, the number of RIS elements, transmit power and quantization bits on the sum AR according to simulations results. Assuming the BS and the active RIS are respectively deployed at (0, 0, 25) and (5, 100, 30) in the simulation. In a semicircle centered at (5, 100, 1.6) with a radius of 5 metres, the users are randomly generated. We generate the AoA and AoD from [0, 2$\pi$] uniformly. For the consistency of the simulation, the generated angles will be fixed. We set other simulation values as follows: the number of users of $ K = 4 $, the number of BS antennas of $ M = 64 $, and the number of active RIS reflecting elements of $ N = 16 $. Noise power is $ \sigma_n^2=-90 $ dBm, dynamic noise is $ \sigma_{\mathrm{v}}^2=-70 $ dBm\cite{9530750}, and Rician factors are $ \varepsilon_k=10, \delta=1 $.  $ P_{\mathrm{T}}=30 $ dBm, $ P_{\mathrm{SW}} = -10 $ dBm and $ P_{\mathrm{DC}} = -5 $ dBm. The large-scale fading coefficient is expressed as $ Pathloss=-30-10\mathcal{T} \log _{10}\left( l \right)  $, where $ l $ denotes the distance of the $ k $-th user to the RIS (or the RIS to the BS). $ \mathcal{T} $ is the path-loss exponent with $ \mathcal{T}_u =2.8 $ for the link of the $ k $-th user to the RIS and $ \mathcal{T}_r =2.8 $ for the link of the RIS to the BS. To mitigate the spatial correlation between antennas, $ d $ is assumed to be $ \lambda/2 $. In addition, we eliminate the influence of randomness by averaging $ 2\times10^4 $ Monte Carlo realizations.

\begin{figure}[t]
	\centering
	\setlength{\abovecaptionskip}{-1ex} 
	\includegraphics[scale=0.48]{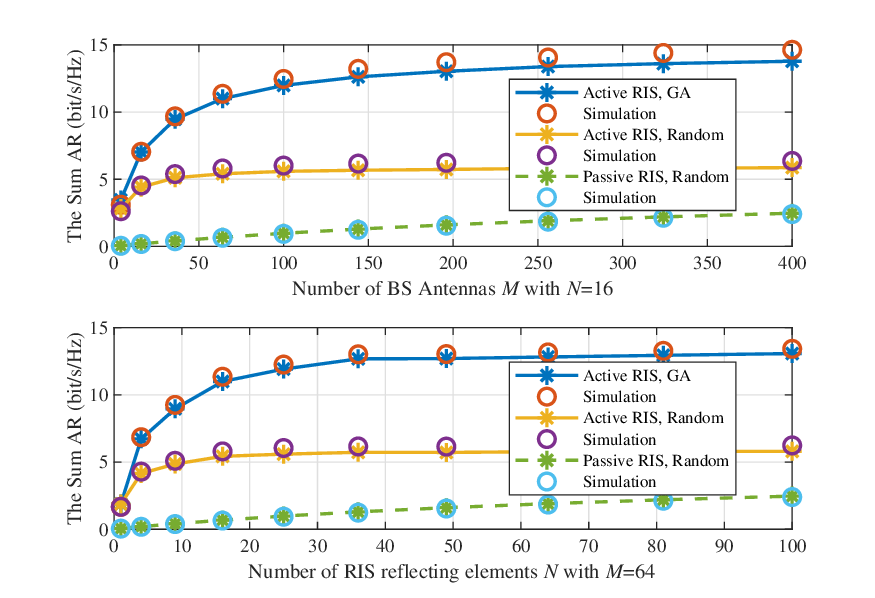}
	\caption{The sum AR versus the number of BS antennas $ M $ and RIS reflecting elements $ N $ with $ b $ = 1.}
	\label{fig2}
	\vspace{-3ex}
\end{figure}
\begin{figure}[t]
	\centering
	\setlength{\abovecaptionskip}{-1ex} 
	\includegraphics[scale=0.48]{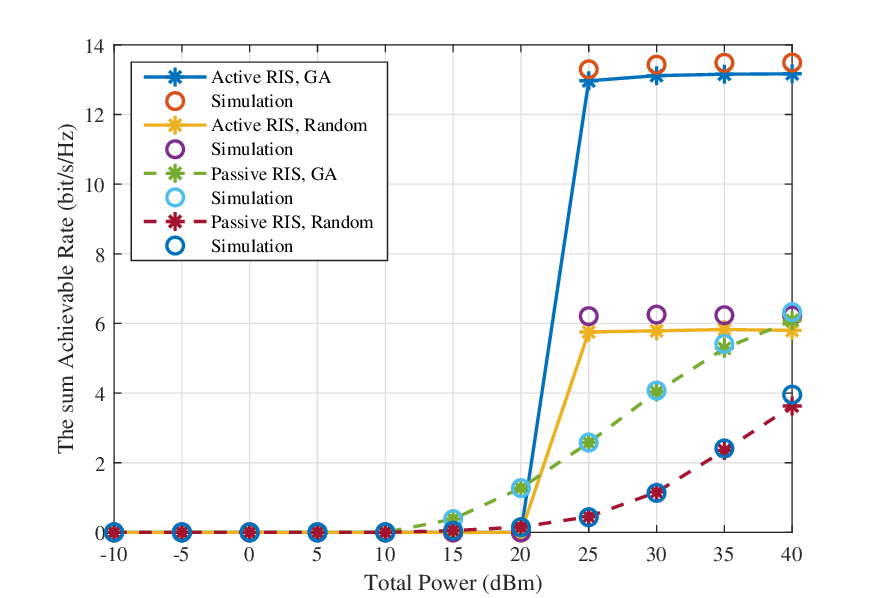}
	\caption{The sum AR versus the total power $ P_{\mathrm{T}} $ with $ b $ = 1.}
	\label{fig3}
		\vspace{-4ex}
\end{figure}
\begin{figure}[t]
	\centering
	\setlength{\abovecaptionskip}{-1ex} 
	\includegraphics[scale=0.475]{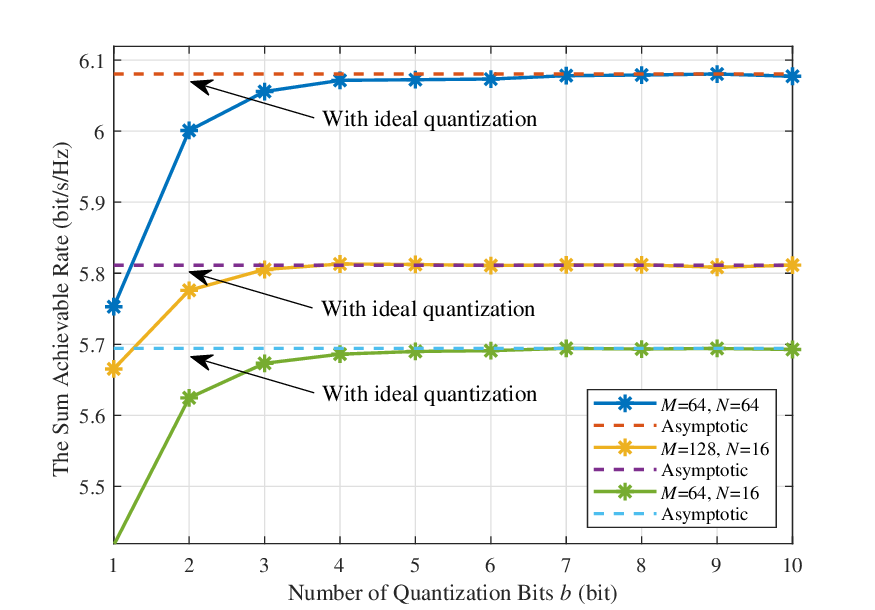}
	\caption{The sum AR versus the number of quantization bits with different $ M $ and $ N $. }
	\label{fig4}
		\vspace{-4ex}
\end{figure}

In Fig. \ref{fig2}, numerical results verify the correctness of the expression in Theorem \ref{thm_appro} and Corollary \ref{c1}, since the ``Active RIS'' and ``Passive RIS'' curves match well with the corresponding ``Simulation'' ones. We can find that the sum AR increases as $ M $ and $ N $ go larger, which provides guiding significance for deploying the RIS in the massive MIMO system. It is also obvious that compared to ``Passive RIS'', ``Active RIS'' is capable of approaching the upper limit of system performance while requiring fewer components such as $ M=64 $ and $ N=16 $. In addition, the system performance can be notablely enhanced by the method in Section \ref{4}.

Based on Fig. \ref{fig3}, we illustrate the impacts of the total power on the sum AR ($ N = 128 $). It is obvious that when $ P_{\mathrm{T}} $ is lower than 20 dBm, the sum AR of ``Passive RIS'' outperforms ``Active RIS''. Because the active RIS system consumes additional power $ P_{\mathrm{SW}} $, it has a higher startup threshold. However, ``Active RIS'' is able to achieve a higher rate under sufficient energy conditions. It means that active RIS can greatly alleviate  ``multiplicative fading'' effect.

Fig. \ref{fig4} depicts the sum AR under the quantization bit $ b $ of ADCs. Apparently, in cases with different $ M $ and $ N $, the sum ARs initially increase fast with $ b $ and then gradually converge to different constants. More importantly, when $ b $ equals 4 bits, the system can achieve the similar performance to the case with high resolution ADCs. This means that the communication system in this letter can apply low-resolution ADCs to balance the system performance with the power consumption and hardware cost.
\vspace{-1ex}
\section{Conclusion}
This letter studied an active RIS-aided massive MIMO system with low-resolution ADCs. We drived the closed-form approximate expression of the sum AR and applied a GA-based method to enhance the system performance. According to the numerical results, we validated the ultimate approximate expression and the algorithm's efficacy. It also revealed that the active RIS has more performance advantages and applying low-resolution ADCs in this system can balance the system performance with the deployment costs.

\begin{figure*}[ht]
	\setcounter{equation}{32}
	\begin{align}\label{33}
		&\mathbb{E} \left\{ \left| \mathrm{g}_{km} \right|^2\left| \mathrm{g}_{im} \right|^2 \right\} =\eta ^4u_ku_i\mathbb{E} \big\{ \sum\nolimits_{c=1}^4{\left| \mathrm{g}_{km}^{c} \right|^2\sum\nolimits_{c=1}^4{\left| \mathrm{g}_{im}^{c} \right|^2}} \big\} +4\eta ^4u_ku_i\big( \mathbb{E} \left\{ \mathrm{Re}\left\{ \mathrm{g}_{km}^{1}\left( \mathrm{g}_{km}^{3} \right) ^{\ast} \right\} \mathrm{Re}\left\{ \mathrm{g}_{im}^{1}\left( \mathrm{g}_{im}^{3} \right) ^{\ast} \right\} \right\}\notag\\
		 &\ \ +\!\mathbb{E}\! \left\{ \mathrm{Re}\!\left\{\! \mathrm{g}_{km}^{1}\!\left( \mathrm{g}_{km}^{3} \right) ^{\ast}\! \right\} \!\mathrm{Re}\!\left\{\! \mathrm{g}_{im}^{2}\!\left( \mathrm{g}_{im}^{4} \right) ^{\ast} \!\right\}\! \right\} \!+\!\mathbb{E}\! \left\{\! \mathrm{Re}\!\left\{\! \mathrm{g}_{km}^{2}\!\left( \mathrm{g}_{km}^{4} \right) ^{\ast} \!\right\}\! \mathrm{Re}\!\left\{\! \mathrm{g}_{im}^{1}\!\left( \mathrm{g}_{im}^{3} \right) ^{\ast} \!\right\}\! \right\} \!+\!\mathbb{E}\! \left\{ \mathrm{Re}\!\left\{\! \mathrm{g}_{km}^{2}\!\left( \mathrm{g}_{km}^{4} \right) ^{\ast} \!\right\} \!\mathrm{Re}\!\left\{ \mathrm{g}_{im}^{2}\!\left( \mathrm{g}_{im}^{4} \right) ^{\ast} \!\right\} \right\} \!\big) \notag
	\end{align}
\vspace{-4ex}
    \begin{align}
		 =\eta ^4u_ku_i\left( \delta \varepsilon _k\left| f_k\left( \mathbf{\Phi } \right) \right|^2+\delta N+\varepsilon _kN+N \right) \left( \delta \varepsilon _i\left| f_i\left( \mathbf{\Phi } \right) \right|^2+\delta N+\varepsilon _iN+N \right)\quad\quad\quad\quad\quad\quad\quad\quad\quad\quad\quad\quad\quad\quad\quad\quad\quad\quad \notag
    \end{align}
\vspace{-4.5ex}
    \begin{align}
		 \!\!\!+2\eta ^4u_ku_i\delta \left( \varepsilon _k\varepsilon _i\mathrm{Re}\left\{ f_{k}^{H}\left( \mathbf{\Phi } \right) f_i\left( \mathbf{\Phi } \right) \mathbf{\bar{h}}_{k}^{H}\mathbf{\bar{h}}_i \right\} +\varepsilon _k\left| f_k\left( \mathbf{\Phi } \right) \right|^2+\varepsilon _i\left| f_i\left( \mathbf{\Phi } \right) \right|^2+N \right),\quad\quad\quad\quad\quad\quad\quad\quad\quad\quad\quad\quad\quad\ 
	\end{align}
\vspace{-4.5ex}
\begin{align}\label{34}
	&\!\mathbb{E} \left\{ \left| \mathrm{g}_{km} \right|^4 \right\}=\eta ^4u_k^2
	\big\{ \!\left( \delta \varepsilon _k\left| f_k\!\left( \mathbf{\Phi } \right) \right|^2 \right) ^2\!+\!4\delta \varepsilon _k\left| f_k\!\left( \mathbf{\Phi } \right) \right|^2\left( N\left( \delta \!+\!\varepsilon _k\!+\!1 \right) \!+\!2 \right) 
	\!+\!2N^2\left( \delta \!+\!\varepsilon _k\!+\!1 \right) ^2\!+\!2N\left( 2\delta \!+\!2\varepsilon _k\!+\!1 \right) \!\big\},\quad\ \ 
\end{align}
\vspace{-9ex}
\begin{align}
\ \notag
\end{align}
\vspace{-1ex}
where $ \mathbf{g}_k=\eta \mathbf{H}_2\mathbf{\Phi h}_k=\eta u_k\small( \sqrt{\delta \varepsilon _k}\mathbf{\bar{H}}_2\mathbf{\Phi \bar{h}}_k+\sqrt{\delta}\mathbf{\bar{H}}_2\mathbf{\Phi \tilde{h}}_k+\sqrt{\varepsilon _k}\mathbf{\tilde{H}}_2\mathbf{\Phi \bar{h}}_k+\mathbf{\tilde{H}}_2\mathbf{\Phi \tilde{h}}_k \small) 
 $, $ \mathrm{g}_{km} $ is the $ m $-th entry of $ \mathbf{g}_k $.
 
	\hrulefill
	\vspace{-3ex}
	\setcounter{equation}{26}
\end{figure*}

\vspace{0ex}
\begin{appendix}
	\section{ }
	The approximation of \eqref{16} has been readily proved in \cite{6816003}. By using  \cite[Theorem 1]{9743440}, $ \mathbb{E} \small\{ \left\| \mathbf{g}_k \right\| ^4 \small\} $, $ \mathbb{E} \small\{ \left| \mathbf{g}_{k}^{H}\mathbf{g}_i \right|^2 \small\} $ and $\mathbb{E} \small\{ \left\| \mathbf{g}_k \right\| ^2 \small\}  $ can be easily obtained. Therefore, the main work that follows is to proof \eqref{19} and \eqref{21}. 
	
	Firstly, we give the derivation of $ \mathbb{E} \small\{ \left\| \mathbf{g}_{k}^{H}\mathbf{H}_2\mathbf{\Phi } \right\| ^2 \small\} $ as 
	\vspace{-1ex}
	\begin{align}\label{27}
		\mathbb{E} \left\{ \left\| \mathbf{g}_{k}^{H}\mathbf{H}_2\mathbf{\Phi } \right\| ^2 \right\}\!=\!\eta ^2\mathbb{E} \left\{ \mathbf{h}_{k}^{H}\!\mathbf{\Phi }^H\!\mathbf{H}_{2}^{H}\!\mathbf{H}_2\!\mathbf{H}_{2}^{H}\!\mathbf{H}_2\!\mathbf{\Phi h}_k \right\}.\ 
	\end{align}
Then, we define $ \mathbf{W}\triangleq\mathbf{H}_{2}^{H}\mathbf{H}_2 $. The matrix $ \mathbf{W} $ satisfies non-central Wishart  distribution $ \mathbf{W}\sim \mathcal{W} _N\left( M,\mathbf{S},\mathbf{\Sigma } \right) $, where $ M $ is the degree of freedom, $ \mathbf{S}=\sqrt{\frac{\beta \delta}{1+\delta}}\mathbf{\bar{H}}_2 $ is the mean matrix and $ \mathbf{\Sigma }=\frac{\beta}{1+\delta}\mathbf{I}_N $ represents the covariance matrix. By approximation, the non-central Wishart matrix can be transformed into a central Wishart matrix \cite{steyn1972approximations}. In addition, the degree of freedom keeps the initial value while the covariance matrix can be $ \mathbf{\bar{\Sigma}}\!=\!\mathbf{\Sigma }+\frac{1}{M}\mathbf{SS}^H\!=\!\frac{\beta}{1+\delta}\left( \mathbf{I}_N+\frac{\delta}{M}\mathbf{\bar{H}}_{2}^{H}\mathbf{\bar{H}}_2 \right) $, i.e. $ \mathbf{W}\sim \mathcal{W} _N\left( M,0,\mathbf{\bar{\Sigma}} \right) $. By applying the results in \cite[Eq.(20)]{RN1} and $ \mathrm{Tr}\left( \mathbf{\bar{\Sigma}} \right)=N\beta $, we have
	\begin{align}\label{28}
		&\mathbb{E} \left\{ \mathbf{H}_{2}^{H}\mathbf{H}_2\mathbf{H}_{2}^{H}\mathbf{H}_2 \right\} =\mathbb{E} \left\{ \mathbf{WW} \right\} = M\mathbf{\bar{\Sigma}}\left( M\mathbf{\bar{\Sigma}}+\mathrm{Tr}\left( \mathbf{\bar{\Sigma}} \right) \right)\quad\  \notag
	\end{align}
\vspace{-4.5ex}
\begin{align}
		= \frac{M^2\beta ^2}{\left( \delta +1 \right) ^2}\big( \mathbf{I}_N+\frac{2\delta}{M}\mathbf{\bar{H}}_{2}^{H}\mathbf{\bar{H}}_2+\frac{\delta ^2}{M^2}\mathbf{\bar{H}}_{2}^{H}\mathbf{\bar{H}}_2\mathbf{\bar{H}}_{2}^{H}\mathbf{\bar{H}}_2 \big)\quad\quad\quad\ \notag
	\end{align}
\vspace{-3ex}
		\begin{align}
		+\frac{MN\beta ^2}{\left( \delta +1 \right) ^2}\big( \mathbf{I}_N+\frac{\delta}{M}\mathbf{\bar{H}}_{2}^{H}\mathbf{\bar{H}}_2 \big).\quad\quad\quad\quad\quad\quad\quad\quad\quad
	\end{align}
\vspace{-1ex}Thus, \eqref{27} can be reformulated as
	\begin{align}\label{29}
		&\mathbb{E} \left\{ \mathbf{h}_{k}^{H}\mathbf{\Phi }^H\mathbf{H}_{2}^{H}\mathbf{H}_2\mathbf{H}_{2}^{H}\mathbf{H}_2\mathbf{\Phi h}_k \right\} =\mathbb{E} \big\{ \frac{M^2\beta ^2}{\left( \delta +1 \right) ^2}\big( \mathbf{h}_{k}^{H}\mathbf{h}_k\quad\quad\quad\notag
	\end{align}
\vspace{-4ex}
\begin{align}
		\ \ \!+\frac{2\delta}{M}\mathbf{h}_{k}^{H}\mathbf{\Phi }^H\mathbf{\bar{H}}_{2}^{H}\mathbf{\bar{H}}_2\mathbf{\Phi h}_k\!+\!\frac{\delta ^2}{M^2}\mathbf{h}_{k}^{H}\mathbf{\Phi }^H\mathbf{\bar{H}}_{2}^{H}\mathbf{\bar{H}}_2\mathbf{\bar{H}}_{2}^{H}\mathbf{\bar{H}}_2\mathbf{\Phi h}_k \!\big) \notag
			\end{align}
		\vspace{-3.5ex}
		\begin{align}
		+\frac{MN\beta ^2}{\left( \delta +1 \right) ^2}\big( \mathbf{h}_{k}^{H}\mathbf{h}_k+\frac{\delta}{M}\mathbf{h}_{k}^{H}\mathbf{\Phi }^H\mathbf{\bar{H}}_{2}^{H}\mathbf{\bar{H}}_2\mathbf{\Phi h}_k \big) \big\},\quad\quad\ 
	\end{align}
	\vspace{0ex}where $ \mathbb{E} \left\{ \mathbf{h}_{k}^{H}\mathbf{h}_k \right\} =N\alpha _k $ and the other terms can be calculated as 
	\begin{align}\label{30}
		\mathbb{E} \left\{ \mathbf{h}_{k}^{H}\mathbf{\Phi }^H\mathbf{\bar{H}}_{2}^{H}\mathbf{\bar{H}}_2\mathbf{\Phi h}_k \right\} =\mathrm{Tr}\left( \mathbb{E} \left\{ \mathbf{h}_{k}^{H}\mathbf{\Phi }^H\mathbf{\bar{H}}_{2}^{H}\mathbf{\bar{H}}_2\mathbf{\Phi h}_k \right\} \right)\quad\quad\notag
	\end{align}
\vspace{-4ex}
		\begin{align}
		\!=\frac{M\alpha _k\varepsilon _k}{\varepsilon _k+1}\mathbb{E} \left\{ \left\| \mathbf{a}_{N}^{H}\left( \phi _{t}^{a},\phi _{t}^{e} \right) \mathbf{\Phi \bar{h}}_k \right\| ^2 \right\} +\frac{MN\alpha _k}{\varepsilon _k+1}\quad\quad\quad\quad\quad\ \notag
	\end{align}
\vspace{-3ex}
\begin{align}
		\!\!\!\!\!=\frac{M\alpha _k}{\varepsilon _k\!+\!1}\!\left( \varepsilon _k\!\left| f_k\!\left( \mathbf{\Phi } \right) \right|^2\!+\!N \right),\quad\quad\quad\quad\quad\quad\quad\quad\quad\quad\quad
	\end{align}
\vspace{-3ex}
	\begin{align}\label{31}
		\!\!\!\!\!\mathbb{E}\! \left\{\! \mathbf{h}_{k}^{H}\!\mathbf{\Phi }^H\!\mathbf{\bar{H}}_{2}^{H}\!\mathbf{\bar{H}}_2\mathbf{\bar{H}}_{2}^{H}\!\mathbf{\bar{H}}_2\mathbf{\Phi h}_k \!\right\}\!\!=\!\!\frac{M^2\!N\!\alpha _k}{\varepsilon _k+1}\!\left( \!\varepsilon _k\!\left| f_k\!\left(\! \mathbf{\Phi }\! \right) \right|^2\!\!+\!\!N \!\right)\!\!.
	\end{align}
Then, \eqref{19} can be obtained by substituting \eqref{29} into \eqref{27}. 

Secondly, $ \mathbb{E} \small\{ \mathbf{g}_{k}^{H}\mathrm{diag}\small( p_k\mathbf{GG}^H+\sigma _{n}^{2}\mathbf{I}_M \small) \mathbf{g}_k \small\} $ can be expanded as
\vspace{0ex}
	\begin{align}\label{32}
		\mathbb{E} \left\{ \mathbf{g}_{k}^{H}\mathrm{diag}\left( p_k\mathbf{GG}^H+\sigma _{n}^{2}\mathbf{I}_M \right) \mathbf{g}_k \right\}\quad\quad\quad\quad\quad\quad\quad\quad\quad\quad\notag
			\end{align}
		\vspace{-3.5ex}
		\begin{align}
		=\mathbb{E} \Big\{ \sum_{m=1}^M{\left| \mathrm{g}_{km} \right|^2\big( p_k\sum_{i=1,i\ne k}^K{\left| \mathrm{g}_{im} \right|^2+}p_k\left| \mathrm{g}_{km} \right|^2+\sigma _{n}^{2} \big)} \Big\} \quad\quad\quad\notag
			\end{align}
		\vspace{-2.5ex}
		\begin{align}
		=p_k\sum_{m=1}^M{\sum_{i=1,i\ne k}^K{\mathbb{E} \left\{ \left| \mathrm{g}_{km} \right|^2\left| \mathrm{g}_{im} \right|^2 \right\}}}+p_k\sum_{m=1}^M{\mathbb{E} \left\{ \left| \mathrm{g}_{km} \right|^4 \right\} }\quad\quad\notag
	\end{align}
\vspace{-2.5ex}
	\begin{align}
		+\sigma _{n}^{2}\sum_{m=1}^M{\mathbb{E} \left\{ \left| \mathrm{g}_{km} \right|^2 \right\}},\quad\quad\quad\quad\quad\quad\quad\quad\quad\quad\quad\quad\ \ 
	\end{align}
where $ \mathbb{E} \small\{ \left| \mathrm{g}_{km} \right|^2 \small\}=\frac{1}{M}\mathbb{E} \small\{ \left\| \mathbf{g}_k \right\| ^2 \small\} $ is obtained by \eqref{20}. In addition, $ \mathbb{E} \small\{ \left| \mathrm{g}_{km} \right|^2\left| \mathrm{g}_{im} \right|^2 \small\} $ and $ \mathbb{E} \small\{ \left| \mathrm{g}_{km} \right|^4 \small\} $ are respectively given in \eqref{33} and \eqref{34} at the beginning of this page. We can calculate \eqref{21} by substituting \eqref{20}, \eqref{33} and \eqref{34} into \eqref{32}. Hence, Theorem \ref{thm_appro} is proved.
\end{appendix}


%
\bibliographystyle{IEEEtran}
\bibliography{Refer}

\end{document}